\documentclass[usegraphicx]{emulateapj}

\newcommand{\hii}{\mbox{H\,{\sc ii}}}

\newcommand{\oiii}{\mbox{[O\,{\sc iii}]}}
\newcommand{\oi}{\mbox{[O\,{\sc i}]}}
\newcommand{\nii}{\mbox{[N\,{\sc ii}]}}
\newcommand{\sii}{\mbox{[S\,{\sc ii}]}}
\newcommand{\hi}{\mbox{H\,{\sc i}}}

\newcommand{\pcmcub}{\mbox{${\rm cm^{-3}}$}}
\newcommand{\kmps}{\mbox{${\rm km\;s^{-1}}$}}
\newcommand{\cmps}{\mbox{${\rm cm\;s^{-1}}$}}

\def\mg2{Mg$_2$}

\def\kms{\relax \ifmmode {\,\rm km\,s}^{-1}\else \,km\,s$^{-1}$\fi}
\def\kkms{\relax \ifmmode {\,\rm{K\,km\,s}}^{-1}\else \,K\,km\,s$^{-1}$\fi}
\def\ha{\relax \ifmmode {\rm H}\alpha\else H$\alpha$\fi}
\def\hb{\relax \ifmmode {\rm H}\beta\else H$\beta$\fi}
\def\hi{\relax \ifmmode {\rm H\,{\sc i}}\else H\,{\sc i}\fi}
\def\hii{\relax \ifmmode {\rm H\,{\sc ii}}\else H\,{\sc ii}\fi}
\def\h2{\relax \ifmmode {\rm H}_2\else H$_2$\fi}
\def\lha{\relax \ifmmode L_{{\rm H}\alpha}\else $L_{{\rm H}\alpha}$\fi}
\def\shi{\relax \ifmmode \sigma_{{\rm HI}}\else $\sigma_{\rm HI}$\fi}   
\def\sh2{\relax \ifmmode \sigma_{{\rm H}_2}\else $\sigma_{{\rm H}_2}$\fi}
\def\degr{\hbox{$^\circ$}}
\def\arcmin{\hbox{$^\prime$}}
\def\arcsec{\hbox{$^{\prime\prime}$}}

\def\fdg{\hbox{$.\!\!^\circ$}}
\def\fs{\hbox{$.\!\!^{\rm s}$}}
\def\farcm{\hbox{$.\mkern-4mu^\prime$}}
\def\farcs{\hbox{$.\!\!^{\prime\prime}$}}
\def\degd#1.#2{ #1\fdg#2 }                 
\def\mind#1.#2{ #1\farcm#2 }               
\def\secd#1.#2{ #1\farcs#2 }               
\def\hhh{\ifmmode {\rm ^h}              
         \else {${\rm ^h}$}
         \fi}
\def\sss{\ifmmode {\rm ^s}              
         \else {${\rm ^s}$}
         \fi}
\def\hms#1h#2m#3s{                      
                  \relax
                  \ifmmode #1^{\rm h}\,#2^{\rm m}\,#3^{\rm s}
                  \else \hbox{$#1^{\rm h}\,#2^{\rm m}\,#3^{\rm s}$}
                  \fi
                 }
\def\dms#1d#2m#3s{                      
                  \relax
                  #1\degr\,#2\arcmin\,#3\arcsec
                 }
\def\hmsd#1h#2m#3.#4s{                  
                      \relax
                      \ifmmode #1^{\rm h}\,#2^{\rm m}\,#3\fs#4
                      \else \hbox{$#1^{\rm h}\,#2^{\rm m}\,#3\fs#4$}
                      \fi
                     }
\def\dmsd#1d#2m#3.#4s{                  
                      \relax
                      #1\degr\,#2\arcmin\,#3\farcs#4
                     }
\def\mag{\relax                          
        \ifmmode ^{\rm m}
        \else $^{\rm m}$
        \fi
       }
\def\magd#1.#2{                          
              \relax
              \ifmmode #1^{\rm m}
                       \hskip-0.55em.\hskip0.22em#2
              \else \hbox{#1$^{\rm m}
                    \hskip-0.55em.\hskip0.22em$#2}
              \fi
             }

\newcommand{\ea}{et al.}

\newcommand{\bfi}{\begin{figure}[htb]} 
\newcommand{\bpfi}{\begin{figure}[p]}

\newcommand{\halpha}{$\rm H\alpha$}
\newcommand{\hbeta}{$\rm H\beta$}

\newcommand{\Sauron}{{\texttt{SAURON}}}
\newcommand{\DensePak}{{\texttt{DensePak}}}

\shorttitle{The circumnuclear region of NGC 7742}
\shortauthors{L.~M.~Mazzuca et al.}

\begin{document}


\title{Minor Merger Origin for the Circumnuclear Starburst in NGC 7742}


\author{L. M. Mazzuca}
\affil{NASA Goddard Space Flight Center, Greenbelt, MD 20771.}

\author{M. Sarzi and J. H. Knapen}
\affil{Centre for Astrophysics Research, University of 
Hertfordshire, Hatfield, Herts AL10 9AB, U.K.}

\author{S. Veilleux and R. Swaters}
\affil{Department of Astronomy, University of Maryland, College Park, MD 20742}

\begin{abstract}
We present an emission-line diagnostic analysis of integral-field
spectroscopic observations that cover the central kiloparsec of NGC
7742. This Sa galaxy hosts a spectacular nuclear starburst ring and
nuclear regions characterized by low-ionization emission. The gas in
the ring rotates in the opposite sense to the stars in the galaxy, suggesting a recent merging or acquisition event. The combination of integral-field measurements for the \ha+\nii\ emission lines from \DensePak\ and the \hb\ and \oiii\ emission from \Sauron\
allow the construction of diagnostic diagrams that highlight the
transition from star formation in the nuclear ring to excitation
by high-velocity shocks or by a central AGN towards the
center. \DensePak\ measurements for the \sii\ line ratio reveal very
low gas densities in the nuclear ring, $N_{\rm e}~<~100 \,\pcmcub$,
characteristic of massive \hii\ regions. Comparison with MAPPINGS~III models for starbursts with low gas densities show that the ring is of roughly solar
metallicity. This suggests that the gas in the nuclear ring originated
in a stellar system capable of substantially enriching the gas
metallicity through sustained star formation. We suggest that NGC~7742 cannibalised a smaller galaxy rich in metal-poor gas, and that  star formation episodes in the ring have since increased the metallicity to its present value.

\end{abstract}

\keywords{galaxies: spiral -- galaxies: structure -- galaxies: starburst -- galaxies: nuclei}

\section{Introduction}
\label{sec:intro}

Line emission from galaxies is produced by gas heated and ionized by
hot OB stars, AGN, or shocks. To help interpret the spectral properties of massive
star forming regions and to recognize the role of the various excitation mechanisms, one can utilize diagnostic diagrams. Veilleux \& Osterbrock (1987; hereafter VO87) introduced a diagnostic analysis that is independent of reddening and is based on the ratios of strong emission lines rather close in wavelength.
Kewley \ea\ (2001; hereafter K01) expand upon the VO87 analysis by
incorporating in their photoionization code the spectral
energy distributions of young stellar clusters predicted by stellar
population synthesis models (Starburst 99, Leitherer \ea\ 1999;
PEGASE, Fioc \& Rocca-Volmerange 1997). The resulting
model grids can discriminate reliably between galaxies hosting an AGN
or a nuclear starburst, and in the case of \hii\ regions, can
constrain the metallicity of the starburst and the strength of the
ionizing radiation relative to the density of hydrogen atoms (i.e. the
ionization parameter, $q$).

NGC~7742 is a face-on SA(r)b LINER galaxy, whose 
morphology is dominated by a prominent nuclear ring with a radius of
10\,arcsec or (assuming $D=22$\,Mpc) 1.0\,kpc (Mazzuca et al. 2006). The ring is lined with \hii\ regions that emit strongly in H$\alpha$ (Pogge \& Eskridge 1993; Knapen et al. 2006) but the \ha\ emission equivalent widths of the \hii\ regions show no evidence for
age gradients, indicating random star formation (Mazzuca et al. 2006). The Palomar survey (Ho et al. 1997) classified the nucleus as a T2/L2 intermediate type, where the emission is representative of a mixture of low-ionization lines (i.e., LINER) and lines associated with star formation (\hii\ region). The galaxy is not obviously interacting at the moment, but may well form a pair with NGC~7743, which is 50\, arcmin (320\,kpc) away but within 50\,\kms\ in radial velocity. De Zeeuw et al. (2002, hereafter deZ02) show that the gas in NGC~7742 is counterrotating with respect to the stars within the entire field of the \Sauron\ observations. This suggests a recent merging or acquisition event.

The presence of a nuclear starburst ring and an active nucleus of composite nature in NGC~7742 makes this object particularly attractive for integral-field spectroscopy, which can capture the superposition of \hii\ region and AGN emission-line features. Using the model-based diagnostic diagrams of VO87 \& K01, we aim to delineate the gas conditions in the nucleus and the nuclear ring. In a second paper (Sarzi et al. 2006; Paper~II) we will extend the analysis presented here to a second nuclear ring galaxy, NGC~4314. Paper~II will also contain a more detailed discussion on both data analysis techniques and physical interpretation.

\section{Observations and Data Reduction}
\label{sec:obs} 
In this study we combined data from two integral-field units (IFUs), namely \DensePak\ (Barden \& Wade 1988) on the WIYN telescope and
\Sauron\ (Bacon et al. 2001) on the William Herschel Telescope (WHT).
The combined dataset provides us with the key emission lines,
\hbeta~$\lambda4861$, \oiii~$\lambda\lambda4959,5007$, \oi~$\lambda6300$,
\nii~$\lambda\lambda6548,6583$, \halpha~$\lambda6563$, and
\sii~$\lambda\lambda6716,6731$, which are necessary to produce the
diagnostic diagrams of VO87 and to constrain the gas density.
The \Sauron\ data are from Falc{\'o}n-Barroso et al. (2006), which covers \hbeta~$\lambda4861$ and \oiii~$\lambda\lambda4959,5007$; the \DensePak\ data (L. M. Mazzuca \ea, in prep.) provides the other lines.
Although both instruments are IFUs with similar fields of view
(33$\times$41 and 30$\times$45 arcsec for \Sauron\ and \DensePak,
respectively) the designs are very different. \Sauron\ is a fully
sampled lenslet array, whereas \DensePak\ consists of 89 functional
fibers (2.812 arcsec diameter) bonded into a $7\times13$ staggered
rectangle with center-to-center fiber spacings of 3.75 and 3.25
arcsec, horizontally and vertically. For NGC~7742, \Sauron\ was configured in low-resolution mode, for a spectral resolution of 4.2\,\AA\ (FWHM, $\sigma_{\rm inst}=108$\,\kmps). \DensePak\ used the \#860 grating centered around \halpha, which yields a spectral resolution of 0.97\,\AA\ (FWHM, $\sigma_{\rm inst}=19$\,\kmps). 

To combine the two datasets, the \DensePak\ spectra were analysed
using the procedure of Sarzi et al. (2006), in order to separate the
stellar absorption and gas emission contribution to the observed
spectra.
To describe the stellar continuum we have used linear combinations of
single stellar population (SSP) models based on the MILES library of
S\'anchez-Bl\'azquez et al. (2006), which will be presented in detail in A.~Vazdekis et al. (in prep.). These templates are particularly suited to match the \DensePak\ data, owing to their high spectral resolution ($\sigma_{\rm inst}=42$\,\kmps), and lead to a very good match to the absorption features, such as the prominent line around 6494\AA\ which is due to Calcium and Iron. Given the
limited number of absorption lines in the \ha+\nii\ region, we have
used only a restricted number of SSP templates adopting a Salpeter
initial mass function, ages of $t=0.1,0.3,1.0,3.0,10$ Gyr, and
metallicities of $Z=0.5,1.0,1.5\, Z_\odot$.
Such an accurate description of the stellar continuum is crucial to
correctly estimate the \ha\ and \sii\ line fluxes towards the center
of NGC~7742, where the strength of the gas emission relative to the
stellar continuum (and corresponding absorption features) decreases
very sharply.
Once the emission-line fluxes were measured, the \DensePak\
integrated-flux field was fitted for each of the three pointings taken
during the run to derive the position of the galaxy center. The
pointings were then shifted to a common central position, adjusted to
the same continuum flux level, which varied for each pointing, and
then merged on a regular grid matching the sampling of the \Sauron\
data (i.e., 0.8 x 0.8 arcsec). The same process was then used to merge the emission-line
fluxes.

 \begin{figure}[ht] 
 \begin{center}
 \includegraphics[width=\columnwidth,bbllx=20pt,bblly=10pt,bburx=510pt,bbury=396pt]{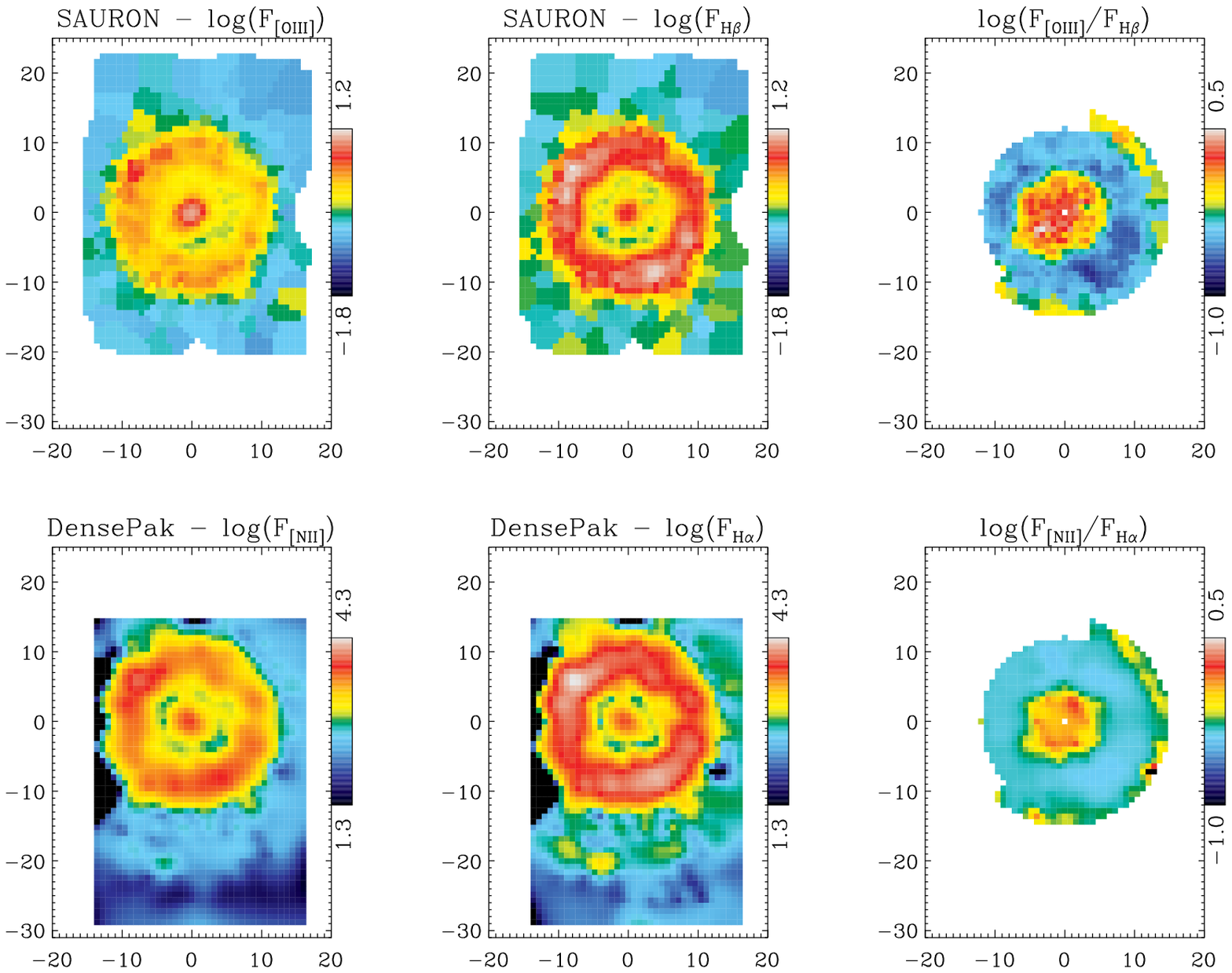}
 \end{center}
 \caption{\Sauron\ and \DensePak\ maps for the \oiii\ and \hb\
 emission (top left and middle panels) and for the \nii\ and
 \ha\ emission (lower left and middle panels), together with maps over
 a restricted common area (to avoid regions where the DensePak coverage is poor) for the \oiii~$\lambda5007$/\hb\ and \nii~$\lambda6583$/\ha\ line ratios (right panels), avoiding regions where the DENSEPAK coverage is poor. Axes are labelled with offsets in RA and DEC (in arcsec), with North up and East to the left.}
 \label{fig:maps} 
 \end{figure} 

\section{Results}
\label{sec:results}

Figure~\ref{fig:maps} shows the reconstructed images for the
distribution of the \ha\ and \nii~$\lambda6583$ emission as measured
with \DensePak, together with similar maps for the \hb\ and
\oiii~$\lambda5007$ fluxes obtained with \Sauron. Despite the
different designs of these two IFUs and the considerably coarser
spatial sampling of \DensePak, the overall morphology of the
recombination and forbidden lines agree very well. The rightmost
panels of Fig.~\ref{fig:maps} show maps for the
\nii~$\lambda6583$/\ha\ and \oiii~$\lambda5007$/\hb\ line ratios,
which will be used to construct the corresponding VO87 diagnostic
diagram separating star formation from other sources of
ionization. 

In the following we will constrain the gas density using the ratio of
the \sii\ doublet lines (\S~\ref{subsec:SIIratios}) and present
\nii/\ha\ vs. \oiii/\hb\ and \nii/\sii\ vs. \oiii/\hb\ diagnostic
diagrams to constrain the role of the various ionizing mechanisms in
different regions of NGC~7742 as well as the metallicity of the
starburst in the circumnuclear regions (\S~\ref{subsec:diagn}).

 \begin{figure}[ht] 
 \epsscale{1.0} 
 \plotone{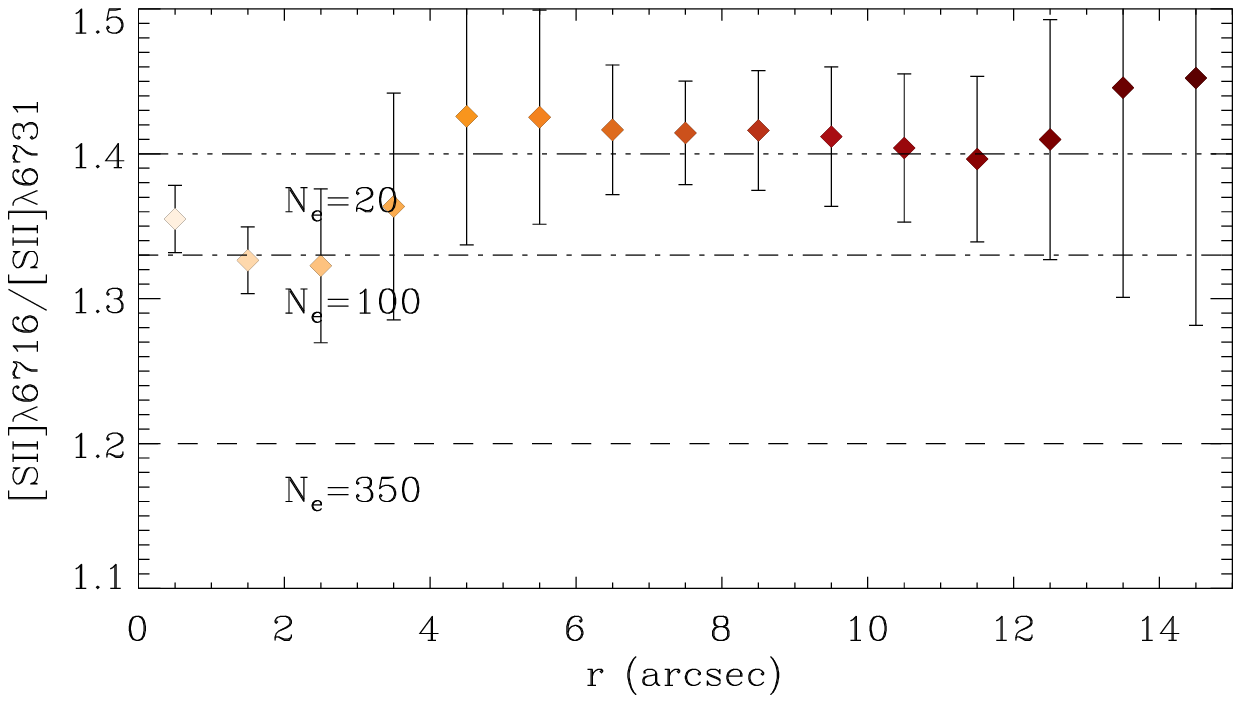}
 \caption{Radial variation of the \sii~$\lambda\lambda6716,6731$ line ratio, which is a tracer of the electron density $N_{\rm e}$. The points show average values of the
 \sii\ line ratio at different radii with error bars indicating the azimuthal variation in this measurement. The horizontal lines indicate the \sii\ line ratio corresponding to specific values of $N_{\rm e}$, assuming gas temperature of $T=10^4$K.}
 \label{fig:SIIratios} 
 \end{figure} 
 
 \subsection{Gas Density}
\label{subsec:SIIratios}

The relative strength of the lines in the \sii\ doublet is a strong
function of the gas electron density $N_{\rm e}$, only depending
weakly on the gas temperature. The observed line ratio can therefore
be used to estimate $N_{\rm e}$, although for low and high values of
$N_{\rm e}$, the \sii~$\lambda6716$/\sii~$\lambda6731$ ratio saturates
to values of $\sim1.4$ and $\sim0.8$, respectively (Osterbrock 1989).
Figure~\ref{fig:SIIratios} shows, as a function of galactic radius,
the ratio of the \sii\ lines within the region of NGC~7742 considered
for the emission-line diagnostic analysis (see
Fig.~\ref{fig:maps}). 
The \sii\ lines ratio indicate, on average, $N_{\rm e}=20
\,\pcmcub$. However, due to the poor sensitivity of this ratio below
$N_{\rm e}=100 \,\pcmcub$, we conclude that the ring is predominantly
populated by clouds of very low electronic density, $N_{\rm e}~<~100
\,\pcmcub$. 
Such low values for $N_{\rm e}$ are typical of extragalactic \hii\ regions, but considerably smaller than observed in the central regions of starburst galaxies, where the typical value is $N_{\rm e}\sim350\,\pcmcub$ within 1 kpc (K01).
$N_{\rm e}$ increases towards the center of NGC~7742, although only to
$N_{\rm e}\sim100\,\pcmcub$. Following this result, we will adopt the lower option of gas densities, $N_{\rm e}=10\,\pcmcub$, for the MAPPINGS~III photoionization models (K01) that will be used in the next section to further interpret our data.

 \begin{figure*}[ht]
 \begin{center}
 \epsscale{1.0}
 \plottwo{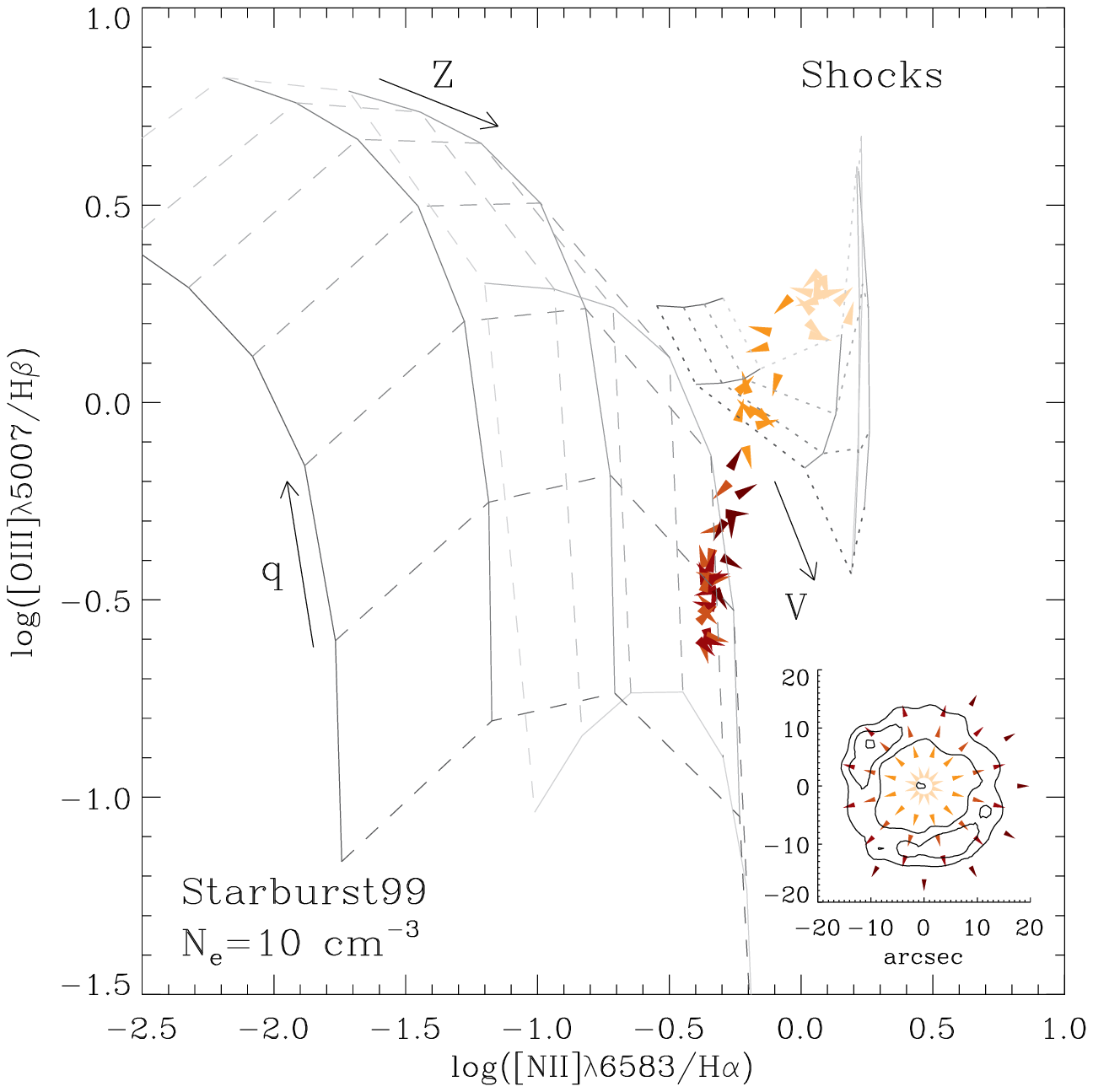}{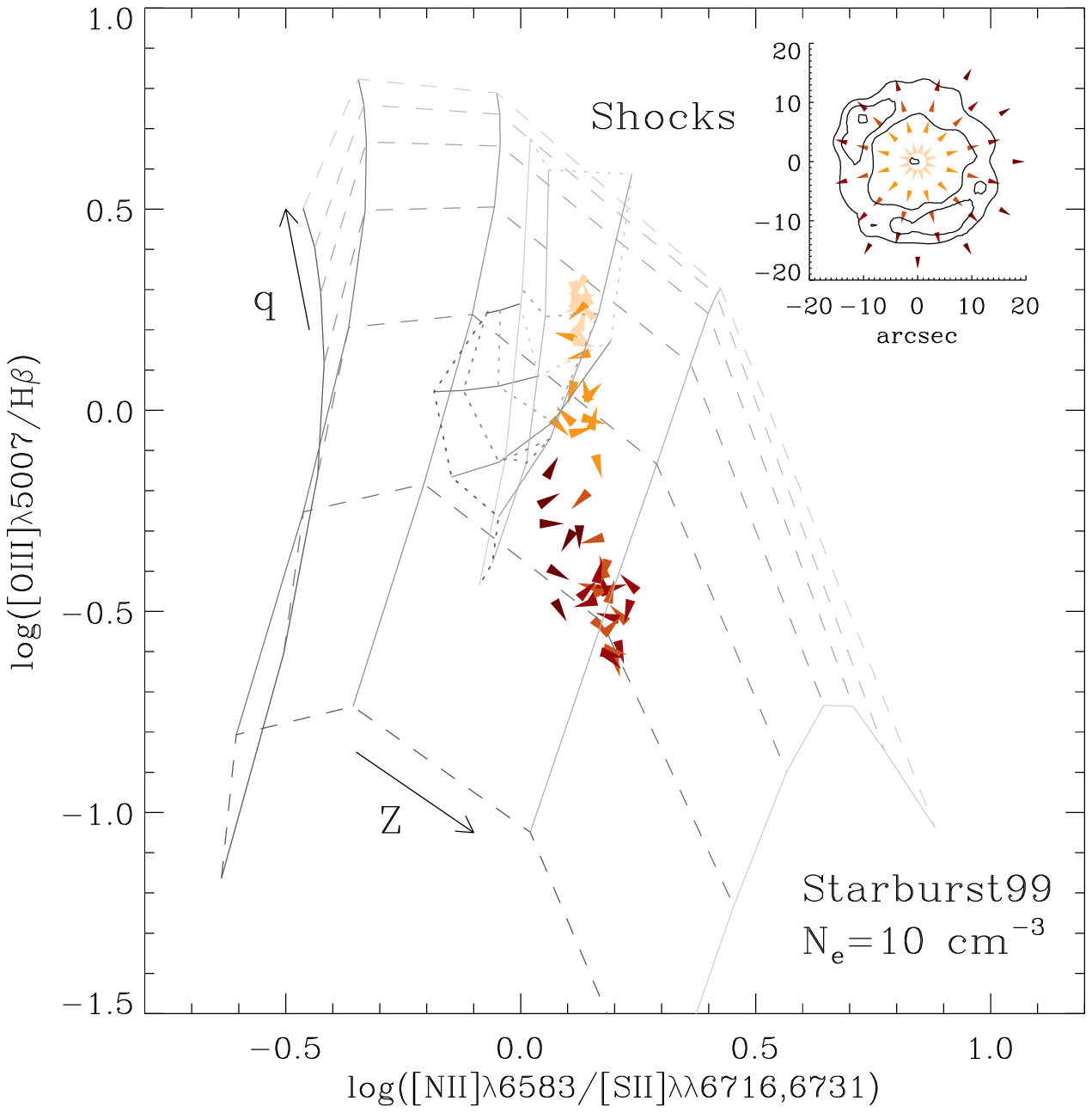}
 \end{center}
 \caption{Diagnostic diagrams of \oiii~$\lambda5007$/\hb vs. \nii$\lambda6583$/\ha (left) and \oiii~$\lambda5007$/\hb vs. \nii$\lambda6583$/\sii~$\lambda\lambda6716,6731$ (right). Data from both \Sauron\
 and \DensePak\ observations are shown by the filled triangles, and
 were extracted in coronal sections at different distances from the
 center and at different position angles. In both panels the inset
 shows the location of these extractions in the galactic frame, with
 the contours showing the \ha\ circumnuclear ring.
 The symbols are color-coded from lighter to darker tones for
 increasing radii and oriented according to the position angle of
 coronal sections apertures. In each panel the prediction of
 MAPPINGS III starburst or shock models are shown by the grid of solid
 and dashed lines or of solid and dotted lines, respectively. For the
 starburst models, which assume a gas density of $N_e\sim10\,\pcmcub$,
 solid lines of increasingly lighter shades of grey show models of
 constant metallicity equal to $Z=0.05, 0.2, 0.4, 1.0, 2.0\,{\rm Z}_\odot$,
 whereas the dashed lines show models with progressively larger values
 for the ionization parameter $q=5e6, 1e7, 2e7, 4e7, 8e7, 1.5e8, 3e8\,\cmps$.
 Similarly, for the shock models (no precursor \hii\ region), the grey solid lines show models with increasing shock velocity $V_{s}=150, 200, 300, 500, 750, 1000\,\kmps$, and the dotted lines models with magnetic parameter $b=0.5,
 1.0, 2.0, 4.0$.}
 \label{fig:diagn}
 \end{figure*}
 
\subsection{Excitation Mechanisms}
\label{subsec:diagn}

Among the reddening-insensitive diagnostic diagrams introduced by
VO87, the \oi~$\lambda6300$/\ha\ vs. \oiii~$\lambda5007$/\hb\
diagnostic is arguably the best suited to separate OB-star excitation from
other excitation sources, since the \oi\ emission originates in
partially ionized regions arising only in the presence of a hard
ionizing spectrum. Unfortunately in NGC~7742 the weak \oi\ emission is detected only in
the nuclear ring, where the equivalent width of the emission line is the largest.
Furthermore, Dopita et al. (2000) showed that the models cannot
distinguish precisely between ionization by hot stars and other
sources of ionization using the \oi/\ha\ and, to some extent, the
\sii/\ha\ diagnostics, as these ratios can be affected by weak shocks
from supernovae.

For all these reasons, we thus use the alternative \nii/\ha\ vs. \oiii/\hb\  diagnostic
diagram, which we show in Fig.~3 (left panel) together with the predictions of MAPPINGS~III models for \hii\ regions in an instantaneous starburst or for clouds excited by shocks.\footnote{Both sets of models were obtained from the MAPPINGS~III portal at {\tt http://www.ifa.hawaii.edu/\~{}kewley/Mappings/\/}}
The starburst models use a spectral energy distribution obtained from
Starburst99 (Leitherer et al. 1999) and assume a range of metallicity
$Z$ for both stars and gas in the starburst and different values of $q$. The shock models assume a range of values for the shock velocity $V_{\rm s}$ and for the magnetic parameter $b$. For a thorough description of these models see Dopita et al. (2000) and Allen, Dopita \& Tsvetanov (1998).

The \nii/\ha\ vs. \oiii/\hb\ diagram (Fig.~3) clearly separates
the emission arising in the nuclear ring (red and orange
triangles) from emitting regions both outside (brown) and inside the
ring (yellow).  The line emission from the ring region is consistent
with the predictions of the starburst models, whereas the position of
the data points measured outside the ring suggests that shocks may play
a role in this region, particularly near the inner edge of the
ring. As far as the nuclear regions are concerned, however, it is
important to keep in mind that models for photoionization by an AGN
can predict line ratios very similar to those shown for the shock
models (Allen et al. 1998), so that excitation by a central engine
cannot be excluded.

Although the \nii/\ha\ vs. \oiii/\hb\ diagnostic diagram is useful to
show the role of different excitation mechanisms in different galactic
regions, it cannot generally be used to derive the physical condition
of the \hii\ regions, except in the very low metallicity regime. The
models in fact fold around, so that any given location in
the diagnostic diagram occupied by starburst models with $Z>0.2
Z_\odot$ could correspond either to an \hii\ region of low
metallicity and low ionization parameter or to an \hii\ region with
considerably larger values of $Z$ and $q$.
In the combined wavelength range of the \DensePak\ and \Sauron\
observations the best diagnostic to constrain the metallicity of the
ring starburst is the \nii/\sii\ ratio. Rubin, Ford \& Whitmore (1984)
were the first to demonstrate the utility of this line ratio, which
serves well to estimate metallicities from slightly sub-solar to
super-solar values (Kewley \& Dopita 2002).
The right panel of Fig.~\ref{fig:diagn} shows the location of our
data in the \nii/\sii\ vs. \oiii/\hb\ diagnostic diagram, which
unequivocally demonstrates that the nuclear ring is
predominantly populated by \hii\ regions with near-solar metallicity.
Having learned from the \nii/\ha\ vs. \oiii/\hb\ diagram that, both
inside and outside the ring, sources other than OB stars may contribute
to the gas excitation, we note that the regions outside the ring
(brown symbols) seem to approach the shock-model grid from a different
direction than the point inside the ring (orange symbols). This
suggests that if indeed shocks are contributing to heating and/or
ionizing the gas, they may occur under different conditions at the
opposite edges of the ring.
\section{Discussion and Conclusions}
We have combined two integral-field spectroscopic data sets to study
the role of various ionizing mechanisms in powering the gas emission
in the central kpc of NGC~7742. Our spatially resolved emission-line
diagnostic analysis reveals a sharp transition between regions powered
by OB stars in the nuclear ring to LINER-like emission both
inside and outside the ring.
Comparison with models for shock ionization indicates that shocks can
contribute to the gas excitation in these regions, in particular
outside the ring. Toward the center photoionization by a central AGN
could also be important.
The ratio of the \sii\ doublet lines indicates low electron density,
$N_{\rm e}<100\,\pcmcub$, in the ring, where comparison with
MAPPINGS~III photoionization models shows that the gas and stars have
approximately solar metallicity.

NGC~7742 shows a clear signature of past galaxy interaction, namely the counter-rotation of the gas and the stars. Although NGC~7743
accompanies NGC~7742 and is sufficiently massive that it could, in
principle, have provided the metal-rich gaseous material observed in the nuclear ring, the large distance between the two galaxies and the morphologically pristine
characteristics of both objects rule out a recent interaction.
This leads us to consider an alternative scenario, whereby NGC~7742
cannibalized a smaller galaxy, rich in gas of low metallicity. This
material was then funneled into the ring where star formation proceeded
long enough for the metallicity of the starburst to increase to its
present value.
Falc{\'o}n-Barroso et al. estimated a mass for the ionised-gas in
NGC~7742 of few $10^6{M}_\odot$, a lower limit to the total amount of gas
in the ring. To reach an oxygen mass fraction of 1\%, corresponding to solar
metallicity, it would take just tens of Myr to have enough supernovae
injecting enriched material in the ring, assuming a star formation rate of few ${M}_\odot{\rm yr}^{-1}$. 

We propose that the gas which has fallen into NGC~7742 has reached a
rather relaxed, though counterrotating, state where massive star
formation in the nuclear ring has gradually enriched the metallicity. An
analysis of the absorption-line indices in the nuclear ring of NGC~7742 by Allard et
al. (2006, in prep.) supports this by confirming that massive star formation has
occurred intermittently there for maybe half a Gyr, in a very similar
way as recently modelled in detail by Allard et al. (2006) for
M100. Like the ring in that galaxy, the one in NGC~7742 is thus a
stable configuration which can easily enrich its metallicity.

All the observational evidence presented or reviewed here thus
suggests strongly that the counterrotating gas came into NGC~7742
through a minor merger event at some time in the history of the
galaxy, and that this gas has since reached solar metallicity as a
result of the massive star formation in the nuclear ring. Although the
counterrotation is unique to the ring host NGC~7742, a very similar
picture of a minor merger at the origin of the nuclear ring in the
non-barred galaxy NGC~278 was proposed on the basis of the disturbed
\hi\ morphology and kinematics in NGC~278 (Knapen et al. 2004). 

More constraints on the gas dynamics are needed to construct a
detailed chemo-dynamical model allowing to verify this
scenario. Alternatively, deeper spectroscopic data could be used to
constrain the metallicity of the gas outside the ring, which according
to our model should have a lower metallicity than the nuclear ring material. In addition, deep \hi\ imaging might reveal the results of the minor merger we postulate for NGC~7742, possibly in the form of tidal tails or disturbed
kinematics. 

\acknowledgments{We thank Joe Shields, John Barnes, and Uta Fritze for
useful discussions. JHK acknowledges support from the Leverhulme
Trust in the form of a Leverhulme Research Fellowship. The WHT is
operated on the island of La Palma by the Isaac Newton Group in the
Spanish Observatorio del Roque de los Muchachos of the Instituto de
Astrof\'\i sica de Canarias. SV is partially supported by NSF/CAREER grant AST~98-74973}

\end{document}